\documentclass{emulateapj}

\usepackage{epsfig}
\usepackage{amsmath}

\newfont{\myfont}{cmmib10}

\newcommand{\btheta}{\hbox{\myfont \symbol{18} }}

\begin{document}


\title{The microarcsecond structure of an AGN jet via interstellar scintillation}


\author{J.-P. Macquart\altaffilmark{1}, L.E.H. Godfrey, H.E. Bignall, J.A.~Hodgson\altaffilmark{2}}
\affil{ICRAR/Curtin University, Bentley, WA 6845, Australia
}

\altaffiltext{1}{ARC Centre of Excellence for All-Sky Astrophysics (CAASTRO)}
\altaffiltext{2}{Now at Max-Planck-Institute f\"ur Radioastronomie, Auf dem H\"ugel 69, 53121, Bonn, Germany}

\begin{abstract}

We describe a new tool for studying the structure and physical characteristics of ultracompact AGN jets and their surroundings with $\mu$as precision.  This tool is based on the frequency dependence of the light curves observed for intra-day variable radio sources, where the variability is caused by interstellar scintillation. We apply this method to PKS 1257--326 to resolve the core-shift as a function of frequency on scales well below $\sim 12\,\mu$as.   We find that the frequency dependence of the position of the scintillating component is  $r \propto \nu^{-0.1 \pm 0.24}$ (99$\%$ confidence interval) and the frequency dependence of the size of the scintillating component is $d \propto \nu^{-0.64 \pm 0.006}$. Together, these results imply that the jet opening angle increases with distance along the jet: $d \propto r^{n_d}$ with $n_d > 1.8$. We show that the flaring of the jet, and flat frequency dependence of the core position is broadly consistent with a model in which the jet is hydrostatically confined and traversing a steep pressure gradient in the confining medium with $p \propto r^{-n_p}$ and $n_p \gtrsim 7$. Such steep pressure gradients have previously been suggested based on VLBI studies of the frequency dependent core shifts in AGN. 
\end{abstract}



\section{Introduction} \label{sec:intro}

The brightest, most compact feature of an AGN jet, the ``core", is identified with the part of the jet at which the optical depth ($\tau_\nu$) is of order unity \citep{blandford79}, and is often referred to as the $\tau_\nu = 1$ surface, or photosphere. Due to positional variation of the opacity in the jet and/or surrounding medium, the position of the $\tau_\nu = 1$ surface is frequency dependent, and therefore, so too is the absolute position of the core \citep[eg.][]{konigl81, lobanov98, kovalev08, sokolovsky11}. The frequency dependent position of the core is referred to simply as the core shift. 

The core shift effect provides an observational tool with which to investigate the structure and physical conditions in parsec-scale AGN jets. Moreover, modelling the effect may provide information about the confinement mechanism and pressure gradients in the external medium. The core shift effect is also relevant to the quest for high precision absolute astrometry for the International Celestial Reference Frame, as it can introduce a significant offset in positions determined using group delay measurements \citep{porcas09}. 

The magnitude of the core shift between 2.3 and 8.4 GHz is typically of order a few hundred $\mu$as or less \citep{osullivan09, sokolovsky11, pushkarev12}, and therefore detecting this effect requires very high accuracy registration of images at two or more frequencies. Despite the technical challenges, the core-shift effect obtained from VLBI imaging has been reported for an ever-increasing number of radio galaxies \citep{marcaide84, lobanov98, kovalev08, osullivan09, sokolovsky11, pushkarev12}. More recently, \citet{kudryavtseva11} have employed an indirect method to measure the core shift effect based on frequency-dependent time lags of flares observed using single-dish data spanning several years. 

The frequency dependence of core position is typically assumed to follow a power-law of the form $r \propto \nu^{-1/k_r}$. In many sources for which core shifts can be measured with VLBI imaging, the absolute core position varies approximately with the inverse of the frequency (i.\,e. $k_r = 1$) \citep{osullivan09, sokolovsky11}. This situation is consistent with the standard model of a conical jet in which the plasma is in a state of equipartition between particle and magnetic energy densities \citep{blandford79}. However, values of $k_r$ much greater than unity are observed in some sources, which may be due to free-free absorption in the immediate vicinity of the jet, or due to rapid changes in pressure in the external medium if hydrostatic confinement is important \citep{lobanov98}. \citet{lobanov98} has shown that while $k_r \sim 1$ at large distances downstream from the black hole, the value of $k_r$ increases towards the jet base. \citet{kudryavtseva11} have shown that the value of $k_r$ is time-dependent, and correlated with flux density. Finally, the pc-scale jet of M87 is observed to deviate from a conical geometry near to the core \citep{asada12}. Further investigation into the frequency dependence of core position is therefore warranted, and highly relevant to the study of ultracompact AGN jets.

Here we present a potentially powerful new method for the study of ultracompact jets in AGN, which enables simultaneous measurement of the core shift effect and jet geometry to very high precision. This new technique uses auto- and cross-correlation analysis of multi-frequency light curves of a rapidly scintillating AGN to measure the frequency dependence of the position and size of the scintillating component. 

In section \ref{sec:obs} we present the observations and data analysis. In Section \ref{sec:jetstructure} we discuss the mathematical formulation of the auto- and cross-correlation analysis, and derive the frequency dependent source position and size for PKS~1257-326. In Section \ref{sec:discussion} we discuss the implications of our findings, and model the jet in terms of a hydrostatically confined jet traversing a steep pressure gradient. Finally, in Section \ref{sec:conclusions} we present our conclusions.

\section{Observations and data calibration}
\label{sec:obs}

PKS\,1257$-$326 was observed at the ATCA for ten hours on 15 January 2011, with two $2$\,GHz bands, centred on frequencies of 5.5 and 9.0 GHz. The output data included all four polarisation products and 2048 spectral channels each 1\,MHz wide in each of the two bands. 
Flagging and calibration of the data were performed using the {\sl Miriad} software package. The ATCA primary calibrator PKS\,1934$-$638 was used to correct the overall flux density scale and the spectral slope. 
In order to solve for the bandpass and to correct gain amplitudes as a function of time and pointing for each antenna, the secondary calibrator PKS\,1255$-$316, only $1^{\circ}$ from PKS\,1257$-$326, was observed for 1 minute approximately every 20 minutes, interleaved with observations of the target source. Phase self-calibration assuming a point source model was performed with a short (10\,s) solution interval. 
After initial calibration the data were split into 128\, MHz sub-bands for further analysis.

At least 98\% of the total flux density of PKS\,1257$-$326 is unresolved with the ATCA, and there is no significant confusion in the field at frequencies above 4\,GHz. Therefore, to obtain the light curves  of the source variations we averaged the real part of the calibrated visibilities over all baselines and frequency points within each 128\,MHz band. The relative stability of the calibration as a function of time and frequency within each 2\,GHz band is estimated to be $\sim 1$\% or better, based on the PKS\,1255$-$316 data. The bandpass is observed to be stable over the duration of our observations in the 5\,GHz band, but there are small frequency- and elevation-dependent gain variations across the higher frequency band, which were corrected with a time-dependent bandpass solution derived from PKS\,1255$-$316. Although the frequency dependence of the primary calibrator PKS\,1934$-$638 is well known, archival data on the secondary calibrator PKS\,1255$-$316 shows it to be variable by up to $\sim 50$\% on timescales of months to years. Hence there is a small uncertainty in the spectral slope correction for the 9\,GHz band, due to the variations with time and pointing, and the fact that the spectrum of PKS 1255-316 is not known a priori.  In any case, the average spectrum of PKS\,1257$-$326 is relatively smooth across the entire range of frequencies, suggesting that the calibration is accurate.  Moreover any residual constant offsets in the flux density scale which may be present between different frequencies have no effect on the cross-correlation analysis presented in this paper.

\begin{figure*}
\epsscale{1}
\plotone{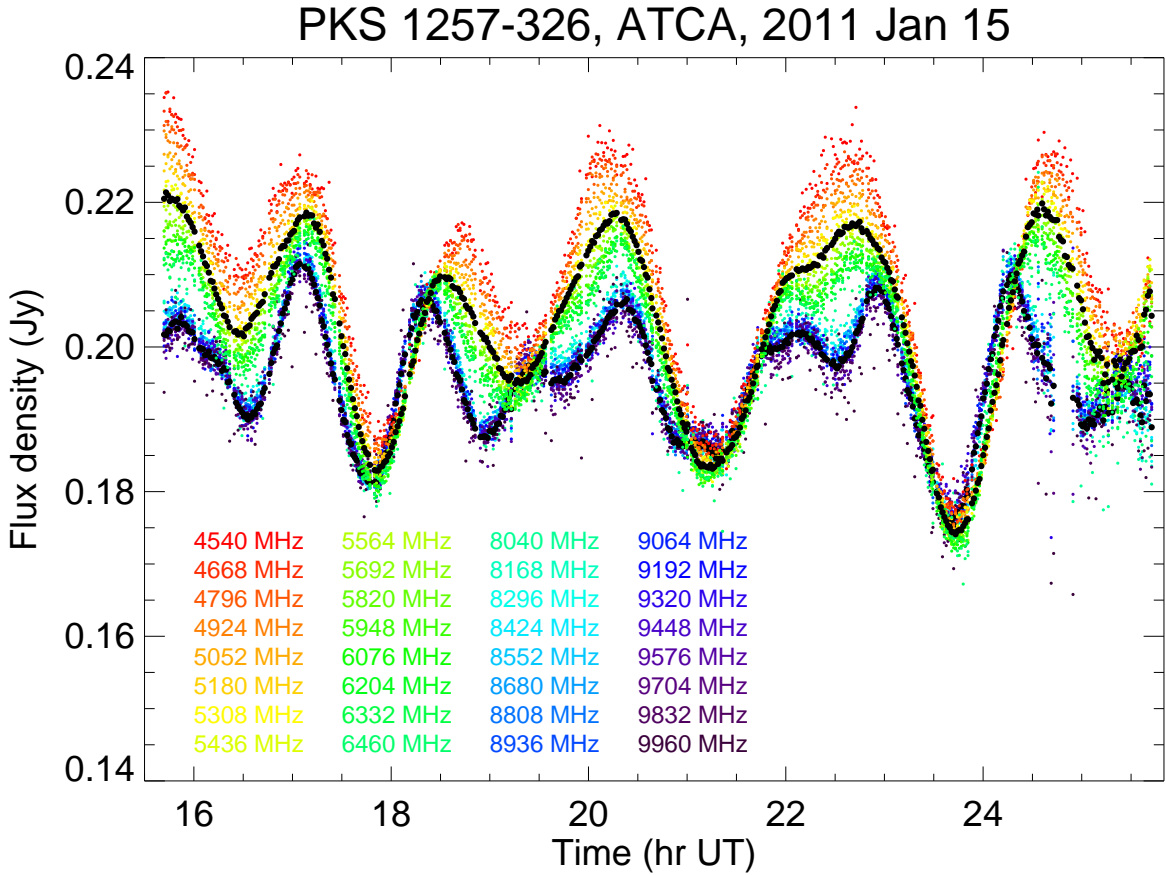}
\caption{PKS\,1257$-$326 flux density measurements as a function of time, plotted with 1-minute averaging, showing each 128\,MHz band in a different colour. Frequency increases from top to bottom. The heavy black points show the average of each 2\,GHz wide-band dataset, corresponding to the centre frequencies of 5.5 and 9 GHz.
\label{fig:lightcurves}} 
\end{figure*}

Figure~\ref{fig:lightcurves} shows the large, rapid intra-hour variations exhibited by PKS\,1257$-$326.


\section{Derivation of Jet Structure} \label{sec:jetstructure}

\subsection{Interpretation}
The rapid fluctuations observed in the centimetre wavelength flux density of the intra-hour variable quasar PKS\,1257$-$326 are due to interstellar scintillation (Bignall et al.\,2003, 2006).  This is established from the measurement of a time delay in the arrival time of the variations between telescopes separated by several thousand kilometers.  The timescale of the variations also undergoes an annual modulation due to relative motion of the Earth about the Sun, which in turn moves relative to the interstellar medium responsible for the variations.  

Inspection of Figure \ref{fig:lightcurves} reveals that there is a time offset in the arrival time of the intensity variations between different frequencies, with the variations at high frequency leading those at lower frequencies.  This behaviour is consistent with observations in previous epochs.  Bignall et al.\,(2003) reported that variations at 4.8 and 8.6\,GHz are closely correlated, and there is a systematic time delay between the variations at these two frequencies.  The magnitude of the time delay was observed to follow an annual cycle which is not identical to the annual cycle in variability timescale but, it was argued, can be explained on the basis of it, as we discuss below. 

PKS 1257-326 was monitored with the ATCA at 4.8 and 8.6\,GHz at 19 epochs between 2001 February and 2002 April. Typical observations were over a 12 hour period, and 6 epochs covered 2 $\times$ 12 hours in 48 hour sessions as part of a multi-source monitoring program. The minimum duration of each light curve is $\sim 10$ times the length of the characteristic timescale. In {\em every} one of these epochs, the time delay between 4.8 and 8.6\,GHz has the same sign, with 8.6\,GHz variations always leading.  Moreover, a clear annual cycle is observed in the two-frequency time delay, with the longest delays being observed from late July to mid-August.  Such an annual cycle is expected for a core-shift which remains stable over the course of the year, and the observed annual cycle is well modelled by such a shift on a scale of order $10 \mu$as (Bignall et al. 2003), although the precise magnitude and direction of the core shift could not be uniquely determined. These data provide strong evidence that the core shift effect dominates over any refractive effects or "jitter" in the ISS pattern.

We argue that the temporal offset in the present data arises as a direct consequence of an angular offset between two compact components within the scintillating source.  The effect may be understood as follows.  When an angular separation, ${\btheta}$, is present between two components, this results in a spatial displacement of their respective scintillation patterns across the plane of an observer by an amount $D \btheta$, whe re $D$ is distance between the observer and the scattering material (Little \& Hewish 1966).  Since the scintillation patterns are in motion across this plane with some velocity ${\bf v}$, the result is a separation in the arrival time of the scintillations associated with each component.  In the present case, a displacement in the lightcurves between closely-spaced frequencies arises because there is an angular offset in the image centroids between the respective frequencies.  For any pair of frequencies, the time delay is, in terms of the centroid offset $\btheta$ (see Appendix \ref{app:Correlation}),
\begin{eqnarray}
\Delta t = - \frac{D \btheta \cdot {\bf v} + (R^2-1) (D \btheta \times \hat{\bf S})({\bf v} \times \hat{\bf S}) }{v^2 + (R^2-1)({\bf v} \times \hat{\bf S})^2}, \label{DeltaTeqn}
\end{eqnarray}
where $R$ is the anisotropy ratio of the scintillation pattern and $\hat{\bf S} = (\cos \beta, \sin \beta)$ is the direction of its major axis, which we measure with respect to the RA axis.   The scintillation parameters have been derived from annual cycle and two-station time-delay measurements, and are given in Table 1.  It is evident that this delay is modulated both by the annual cycle in the magnitude of ${\bf v}$ and by changes in the angle of the velocity vector with respect to ${\btheta}$; this latter effect causes the annual cycle experienced by $\Delta t$ to differ from the annual cycle in scintillation velocity.


Phase gradients in the ISM may, in principle, also cause temporal offsets of lightcurves as a function of frequency in a scintillating source.  However, the offset observed here is difficult to attribute to such an extrinsic cause for several reasons: (i) the sense and magnitude of the delay is constant throughout the dataset; upon dividing the dataset in two halves (in time) and deriving time offsets based on these two halves separately, we find the same offsets to within the margin of error of the estimates.  (ii) The delay is observed over a timescale of 10 hours, whereas refractive phase gradients in the ISM in the regime of weak scintillation for a Kolmogorov spectrum of phase inhomogeneities would occur on the timescale associated with the scintillations, and the time offset should converge to zero as the average is performed over an increasing number of scintles.  Any small jitter in the offset between individual scintles appears to be dominated by the systematic offset.  (iii)  An annual cycle in the time offset is reported by Bignall et al. (2003), indicating that the offset persists on a timescale of greater than a year.



\begin{deluxetable*}{lccc}  \label{tab:scintParams}
\tabletypesize{\scriptsize}
\tablecaption{The parameters of the scintillations in PKS\,1257$-$326.} 
\tablewidth{0pt}
\tablehead{
\colhead{Parameter\tablenotemark{a}} & \colhead{Symbol} & \colhead{B06 value} & \colhead{W09 value} 
}
\startdata
screen distance & $D$ & $10\,$pc & $10\,$pc\\ 
scintillation velocity, ${\bf v}$,  & $v_\alpha $ & $54.12\,$km\,s$^{-1}$ & $33.7\,$km\,s$^{-1}$ \\
\quad on observation date & $v_\delta$ & $-0.05\,$km\,s$^{-1}$ & $0.3\,$km\,s$^{-1}$ \\
anisotropy ratio & $R$ & 12 & 6 \\ 
anisotropy orientiation & $\beta$ & $-35^\circ$ & $-33^\circ$
\enddata
\tablenotetext{a}{Parameters are derived on the basis of the annual cycle in the variability timescale (Bignall et al.\,2003) and measurements of the time delay observed between two stations (Bignall et al.\,2006, henceforth B06).  An alternate fit to the scintillation data provided by Walker, de Bruyn \& Bignall (2009) (henceforth W09) is also listed. The scintillation velocity quoted here is the addition of Earth's relative to the Sun velocity on the observation date and the velocity of the scattering medium relative to the Sun.}
\end{deluxetable*} \label{tab:scintParams}

\subsection{Time delay measurement} \label{subsec:tdelay}

\begin{figure}
\epsscale{1}
\plotone{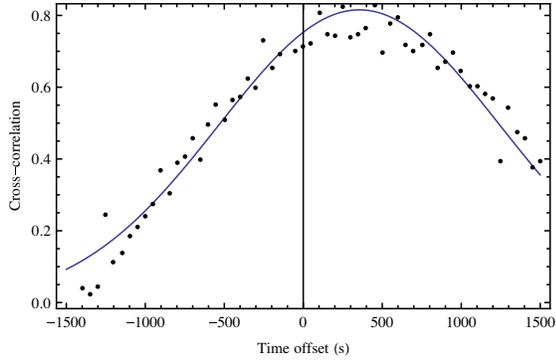}
\caption{The cross correlation between the lightcurves at 8040 and 4540\,MHz, and the associated best-fit gaussian. The peak of the gaussian represents the location of the time-delay.  The positive offset of the peak indicates that the variations at 8040\,MHz lead those at lower frequency. \label{fig:CrossCorr}} 
\end{figure}
\begin{figure}
\epsscale{1.2}
\plotone{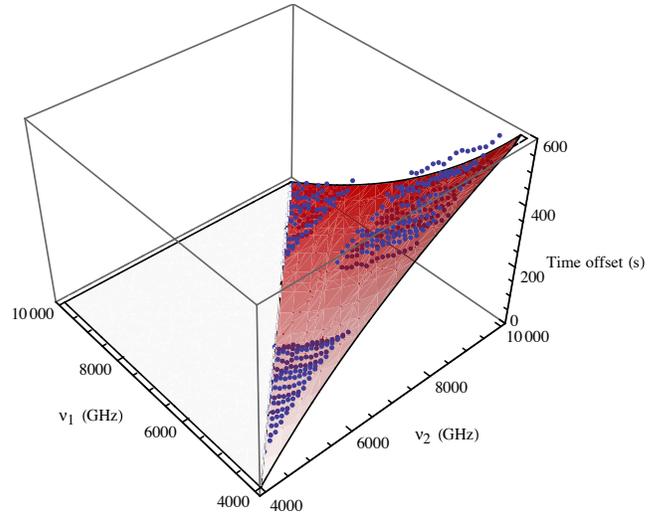}
\caption{The time delay associated with each pair of lightcurves measured at frequencies $\nu_1$ and $\nu_2$.  The red surface corresponds to the best-fitting model of the form $\Delta t = A (\nu_1^{-\zeta} - \nu_2^{-\zeta})$ (see text for details).} \label{fig:DelaySurface}
\end{figure}

To determine the relative time delay between each pair of lightcurves,  $I (t,\nu_1)$ and $I(t,\nu_2)$, we computed the cross-correlation function,
\begin{eqnarray}
C(\Delta t; \nu_1,\nu_2) = \frac{\langle [I(t',\nu_1) - \bar{I}(t,\nu_1)][I(t'+\Delta t,\nu_2) - \bar{I}(t,\nu_2)] \rangle }{ \sqrt{{\rm var}[I(t,\nu_1)] {\rm var}[I(t,\nu_2)] } }. \nonumber \\
\end{eqnarray}
A peak in the cross correlation at positive delay, $\Delta t$, indicates that the fluctuations at frequency $\nu_2$ precede those at $\nu_1$.  We fitted a gaussian of the form,
\begin{eqnarray}
C(t) = A \exp \left[ -\frac{(t - t_0)^2}{B^2} \right], \label{fitform}
\end{eqnarray}
to the inner part of the cross-correlation function, $C(\Delta t;\nu_1,\nu_1)$ (equivalent to the auto-correlation function), to obtain an estimate of the time delay between each frequency-lightcurve pair and its associated error.  An example cross-correlation function and its associated fit is shown in Fig.\,\ref{fig:CrossCorr}.  
Typical errors in the estimated delay are 50\,s.  The derived delays as a function of $\nu_1$ and $\nu_2$ are shown in Fig.\,\ref{fig:DelaySurface}.  The estimated 50\,s uncertainties in the delay between each frequency pair are derived from least-squares fitting of a gaussian to the peak of the delay, thus the errors are directly related to the width of the peak in the cross-correlation functions\footnote{This error is significantly smaller than the time delay errors that might be estimated on the basis of the data presented in Bignall et al.\,(2003). ÊAs is evident from Fig. 2 in Bignall et al. (2003), the data in most of those epochs contain considerably few scintles from which to estimate the time lag. ÊThus one would expect the errors to be larger relative to the 2011 data. ÊThe dataset obtained on 4 Jan 2001 contains a comparable number of scintles, but here the temporal sampling of the lightcurve was too sparse to estimate a time delay.}.  These uncertainties are in turn derived from the least squares fit to the cross-correlation function; the errors in the individual points in the cross-correlation function are dominated, at low time lags, by Poisson errors associated with the number of independent measurements of the cross-correlation measureable from the lightcurve.  This error is indicated by the scatter between points in the CCF.  Fitting to the CCF yields a typical formal error of  $\approx 20\,$s.  However the high degree of cross correlation between the lightcurves in our observations means that the error in the time delay estimate for any given frequency pair is not completely independent from the time delay estimate of any other frequency pair.  It is necessary to take the cross-correlation into account because it becomes an important factor when assessing the significance of the fit to the frequency dependence of the time delay using a least-squares approach (see below), which is the primary reason to estimate the error in the time lag.  Failure to take into account this interdependence in time lag estimates would lead to an overestimate of the significance of the fit to the frequency dependence of the time shift.  It was found empirically, from examination of the reduced chi-squared in the fit procedure that this cross-correlation is taken into account in the error analysis with errors that are $2.5$ times larger than the formal error.

In most models of jet structure the angular offset between the jet base and the centroid of the emission at a frequency $\nu$ is fitted to a power law, $\Delta \theta = A \nu^{-\zeta}$, where $A$ and $\zeta$ are constants to be determined.  Since the time delay is linearly proportional to $| \btheta |$, we fitted a function of the form $\Delta t (\nu_1,\nu_2)  = K (\nu_1^{-\zeta} - \nu_2^{-\zeta})$ to the delays.  The best-fitting parameters are $\zeta = 0.10$ and $K=(1.6 \pm 0.5) \times 10^4\,$s\,MHz$^{\zeta}$, with the 99\% confidence interval of $\zeta$ extending over the range $[-0.14,0.34]$; the best-fitting surface is plotted over the data in Fig.\,\ref{fig:DelaySurface}\footnote{The 99\% confidence limits are calculated, in the conventional way, from the change in the sum of the squares of the residuals of the fit.  Now, in practice, since the lightcurves are correlated between frequencies, the time lag estimate between adjacent frequency pairs are strictly not independent, and this influences the chi-squared estimate.  However, this effect is treated by accounting for the cross-correlation in the estimates of the time delays for each frequency pair.}.  The best-fitting exponent indicates a core-shift dependence on frequency that is much shallower than the value $\zeta =1$ typically found in other quasars on the basis of VLBI measurements \citep[e.g.][]{osullivan09, sokolovsky11}.
 
\begin{figure}
\epsscale{1}
\plotone{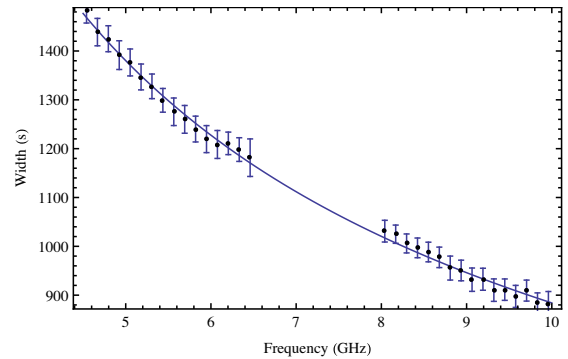}
\caption{The timescale of the scintillations as a function of frequency.  The line represents the best-fitting model of the form $\tau = K \nu^{-\beta}$ (see text for details).  Note that the intrinsic scatter between adjacent points is smaller than the error bars: this is because the lightcurves are highly correlated leading to a smaller scatter than is reflective of the errors.} \label{fig:SizeScaling}
\end{figure}

We also measure the timescale of the scintillations, which is derived from the parameter $B$ in eq.\,(\ref{fitform}) (and with $\nu_1 = \nu_2$).  Figure\,\ref{fig:SizeScaling} shows that the timescale is well-fit by a relation scaling as $\nu^{-0.642 \pm 0.006}$; a fit using a broken power law reveals the both the 4-5 and 8-9\,GHz timescales follow the same scaling with frequency within the (small) margin of error.  

In the regime in which the angular size of the source, $\theta_{\rm src}$, exceeds the angular size of the Fresnel scale at the distance of the scattering screen, $\theta_{\rm F} = \sqrt{c/2 \pi \nu D}$, the scintillation timescale is linearly proportional to the size of the scintillating component.  In the opposite regime, $\theta_{\rm src} < \theta_{\rm F}$, the scintillation timescale is determined by $D \theta_{\rm F}/v = r_{\rm F}/v$, which scales as $\nu^{-0.5}$.

We wish to determine whether the timescale of the scintillations measures $\theta_{\rm F}$ or $\theta_{\rm src}$.  The expected Fresnel crossing timescale is $1.0 \times 10^3\,(D/10\,{\rm pc})^{1/2} (\nu/5\,{\rm GHz})^{-1/2} (v/{\rm 54\,km\,s}^{-1})^{-1}\,$s.  The observed scintillation timescale at 5.0\,GHz is $1.4 \times 10^3\,$s, so if the source is unresolved there must be an error in the nominal scintillation parameters; if the scintillation velocity is held at its nominal value one must have $D > 19\,$pc, or if the screen distance is held at its nominal value one must have $v<40\,$km\,s$^{-1}$.  The latter option is viable for the scintillation velocity derived by Walker, de Bruyn \& Bignall (2009), so we conclude that it is plausible that the source is unresolved by the scintillations.  However, the fact that the scaling of scintillation timescale is significantly different from $\nu^{-0.5}$ suggests that the source is at least partially resolved in any case.  The smooth trend evident in both timescale and scintillating amplitude with frequency suggests that the source remains resolved over the entire frequency range 5-10\,GHz.


Although tangential to the objectives of this paper, we can, in addition, estimate the amplitude of the scintillating component of the source.  For an extended source of size $\theta_{\rm src}$ in the regime of weak scintillation caused by a Kolmogorov spectrum of turbulent fluctuations, the observed rms, $\langle \delta I^2 \rangle^{1/2}$ can be expressed in the form (Narayan 1992),
\begin{eqnarray}
\frac{\langle \delta I^2 \rangle^{1/2}}{\langle I \rangle} \approx \left( \frac{r_{\rm diff} }{r_{\rm F}}\right)^{5/6} \left\{ 
\begin{array}{ll} 
\left( \frac{\theta_{\rm F}}{\theta_{\rm src}} \right)^{7/6} & \theta_{\rm src} > \theta_{\rm F} \\
1 & \theta_{\rm src} < \theta_{\rm F}, \\
\end{array} \right.  \label{SscintEq}
\end{eqnarray}
where $\langle I \rangle$ is the flux density of the scintillating component of the source and $r_{\rm diff} \propto \nu^{6/5}$ is the diffractive scale, which is determined by the properties of the interstellar turbulence (see Narayan 1992).  In the regime of intermediate scattering one has $r_{\rm diff} \approx r_{\rm F}$, with equality at the transition frequency, which likely occurs in the range 3-7\,GHz based upon the modelling of Walker (1998).  One then solves for $\langle I \rangle$ using the measurements of $\langle \delta I^2 \rangle^{1/2}$. Assuming that the source size exceeds the Fresnel angular scale, one determines $\theta_{\rm src}/\theta_{\rm F}$ from the ratio $t_{\rm scint}/t_{\rm F}$.  We performed a fit using the measured values of $\langle \delta I^2 \rangle$ and $t_{\rm scint}$, and subject to the assumption that $r_{\rm diff} = r_{\rm F}$ at 4\,GHz, to derive a rough estimate of the flux density of the scintillating component:
\begin{eqnarray}
S_{\rm scint} = 19 \, \left( \frac{\nu}{5\,{\rm GHz}} \right)^{0.5} {\rm mJy}.
\end{eqnarray}
We caution that the spectral index of the component derived here is only approximate because the expression for the modulation index in eq.(\ref{SscintEq}) is only approximate at frequencies where $r_{\rm diff} \approx r_{\rm F}$.  A more sophisticated estimate would employ a more complicated approximation to the modulation index near the transition frequency and take into account the anisotropy of the scintillations.

\subsection{Overall source scale}

Although the frequency scaling of the core position and size are the primary results of our analysis, it is possible to roughly relate these measurements back to physical scales within the source.

The Schwarzschild radius of the black hole at the centre of PKS\,1257$-$326 is $R_S = 2. 95 \times 10^{12}$\,M$_9$\,m, where M$_9 = 10^9\,$M$_\odot$ is the BH mass, which is estimated from measurements of the broad line emission width to be $10^{8.8}$\,M$_\odot$ (D'Elia, Padovani \& Landt, 2003).  At the redshift of PKS\,1257$-$326, $z=1.256$, $1\,\mu$as subtends $8.5 \times 10^{-3}$\,pc, equivalent to $89/M_9$ Schwarzschild radii\footnote{We use $H_0=70\,$km\,s$^{-1}$\,Mpc$^{-1}$, $\Omega_m=0.27$ and $\Omega_\Lambda = 0.73$ throughout.}.

The time delay is related to angular structure in the source using eq.\,(\ref{DeltaTeqn}) and the scintillation parameters in Table 1.  This also depends on the angle, $\xi$, that $\btheta$ makes with the right ascension axis, which is unknown.  For a time offset $\Delta t$, the expected amplitude of the angular offset is given by
\begin{eqnarray}
| \btheta | &=& \Delta t \left\{
\begin{array}{lr}
(27.7 \cos \xi + 38.8 \sin \xi)^{-1} \mu{\rm as}, & {\rm B06}, \\
(43.9 \cos \xi + 61.8 \sin \xi)^{-1} \mu{\rm as}, & {\rm W09}, \\
\end{array}
\right.
\end{eqnarray} 
where the two solutions denote the scintillation parameters found by Bignall et al.\,(2006) (B06) and Walker et al.\,(2009) (W09).
For instance, the angular separation implied by the 520\,s delay observed between the lowest (4540\,MHz) and highest (9960\,MHz) frequency bands of our observations is $19\,\mu$as ($12\,\mu$as) for the scintillation parameters of B06 (W09) if $\xi=0$ (i.e.\,if the angular separation is aligned parallel to the right ascension axis).  This translates to a physical scale of 0.16\,pc (0.10\,pc) at the source.  

However, these estimates are subject to uncertainties in both the screen distance and the orientation of the separation $\xi$.  For instance, doubling $D$ above its nominal value of $10\,$pc would result in estimates of $| \btheta |$ lower by a factor of two.  They thus serve only as a guide to the order of magnitude of scales which are probed by these measurements.

In the same vein, we also estimate the angular scale of the source from the variability timescale, using eq.\,(\ref{tscintEq}) in Appendix \ref{app:Correlation} to find the characteristic angular scale of the scintillation pattern,
\begin{eqnarray}
\theta_{\rm src}(\nu)  = \left[ \begin{array}{c} 28 \\ 64 \end{array} \right]  \left( \frac{\nu}{9.96\,{\rm GHz}} \right)^{-0.64}  \left( \frac{D}{10\,{\rm pc}} \right)^{-1} \, \mu{\rm as} \,\, \left[ \begin{array}{c} B06 \\ W09 \end{array} \right].
\end{eqnarray}
The minor axis of the scintillation pattern has a size $\theta_{\rm src}/\sqrt{R}$ while the major axis has a characteristic size of $\theta_{\rm src} \sqrt{R}$.

\section{Discussion} \label{sec:discussion}

\subsection{Observational Constraints}

\begin{figure}[!ht]
\epsscale{0.8}
\plotone{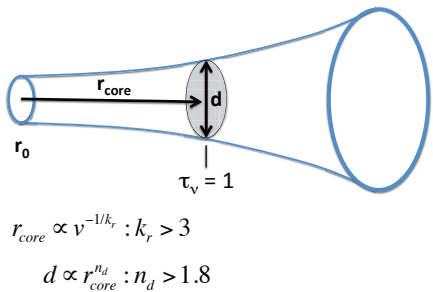}
\caption{A cartoon illustrating the constraints implied by our analysis.} \label{fig:1257_cartoon}
\end{figure}

The foregoing analysis indicates that the centroid of the scintillating component has a frequency dependence of the form: $r_{\rm core} \propto \nu^{-0.1 \pm 0.24}$. In the terminology of \citet{konigl81} and \citet{lobanov98}, $r_{\rm core} \propto \nu^{-1/k_r}$, we have $k_r > 3$. We proceed under the assumption that the source is resolved, as suggested by the frequency dependence of the scintillation time-scale (Section \ref{subsec:tdelay}). In that case, the jet diameter has a frequency dependence of the form: $d \propto \nu^{-0.64 \pm 0.006}$. Taken together, these results imply $d \propto r_{\rm core}^{n_d}$ with $n_d > 1.8$. The scenario implied by our analysis is illustrated in Figure \ref{fig:1257_cartoon}. This is in contrast to the frequency dependence of the core position in VLBI studies: typically $r_{\rm core } \propto \nu^{-1}$, consistent with expectations for a conical jet in which the particle and magnetic energy densities are in equipartition \citep{sokolovsky11}. 

\subsection{Interpretation in terms of hydrostatic jet confinement in a steep pressure gradient}

\citet{lobanov98} suggested that the changes in jet opacity observed for a sample of sources appear consistent with a self-absorbed jet propagating through a region with a steep pressure gradient of the form $p \propto r^{-6}$. Here we consider such a model to explain the inferred frequency dependence of the core size and position in PKS~1257--326. Specifically, we consider whether such a model can simultaneously account for the increase in opening angle along the jet ($d \propto r_{\rm core}^{n_d}$ with $n_d > 1.8$), and the flat frequency dependence of core position ($r_{\rm core} \propto \nu^{-1/k_r}$, with $k_r > 3$). The relevant model equations, for a pressure profile $p \propto r^{-n_p}$, are (see Appendix \ref{app:hydrostatic_confinement})
\begin{eqnarray} \label{eqn:relationships}
n_d &=& \frac{n_p}{4} \nonumber \\ 
k_r &=& \frac{(n_p/4)[(1.5+\alpha)(2+m_B) - 1]  - (1.5 + \alpha) n_\delta}{2.5 + \alpha}  \nonumber \\
\alpha_{\rm core} &=& \frac{5}{2} - \frac{1}{2 k_r} \left[  n_p\left( 1 + \frac{m_B}{4}  \right)  + n_\delta \right] \nonumber
\end{eqnarray}
Here $m_B$ is a parameter determined by the magnetic field geometry as discussed in Appendix B, and $\alpha$ is the optically thin spectral index which directly relates to the electron energy distribution $N(\gamma) \propto \gamma^{-(2 \alpha + 1)}$, as opposed to $\alpha_{\rm core}$, which is the optically thick spectral index of the self-absorbed core. We assume $\alpha = 0.5$, $m_B = 1$ (i.\,e. a predominantly toroidal magnetic field configuration) and take $n_p = 7.2$ so that $n_d = 1.8$. In that case, the jet Lorentz factor goes as $\Gamma / \Gamma_0 = \left( r / r_0 \right)^{1.8}$, so that for reasonable values of the jet viewing angle ($5^\circ < \phi < 40^\circ$) and initial Lorentz factor ($2 \lesssim \Gamma_0 \lesssim 10$), we have $-0.5  \lesssim n_\delta \lesssim 0.5$, which implies:
\begin{eqnarray}
2.7 \lesssim &k_r& \lesssim 3.3, \\ 
0.7 \lesssim &\alpha_{\rm core}& \lesssim 1.2. \nonumber
\end{eqnarray}


There is excellent agreement between the predicted and observed frequency dependence of core position and radial dependence of core size.   The agreement between the predicted and observed core spectral index is poorer, but this is perhaps unsurprising since, as remarked above, observational determination of the core spectral index is marred by uncertainties in the model used to derive it.   
We therefore suggest that this simple model of hydrostatic confinement is broadly consistent with the data, provided $n_p \gtrsim 7.2$. The required pressure gradient in the external medium ($n_p \gtrsim 7.2$) appears very steep, however, such pressure gradients have previously been suggested in studies of core shifts in AGN \citep{lobanov98}. 

Finally, we note that the frequency dependence of core position may also be influenced by free-free absorption in the immediate environment of the jet \citep[e.\,g.][]{lobanov98}. However, we find no evidence for a significant rotation measure, and such an interpretation cannot account for the flaring jet geometry, so that an additional mechanism, such as entrainment, would also be required to explain these results.

\section{Conclusions} \label{sec:conclusions}

We have developed a new technique to probe the structure of AGN jets on micro-arcsecond scales, by using interstellar scintillation to simultaneously determine the shift in the position of the AGN core as a function of frequency, and the frequency scaling of the core size, to high precision.  This approach is amenable to sources which harbour compact ($\la 100\,\mu$as) features. 

This method was applied to broadband observations of PKS\,1257--326 with the Australia Telescope Compact Array.  The scaling of the core-shift is found to be remarkably shallow with frequency; the best fit to the position of the scintillating component in the source scales as $r \propto \nu^{-0.10}$, with the 99\% confidence interval of the index extending over the range $[-0.34, 0.14]$.  This constrasts with previous VLBI studies which typically -- though not exclusively -- find $r \propto \nu^{-1}$.   The scaling of the jet size is also determined, based on the scaling of the scintillation timescale with frequency.  This shows that the jet size scales formally as $d \propto \nu^{-0.64 \pm 0.006}$.  It is possible to determine the scaling of the core shift and jet diameter to high precision because they do not depend critically on complete knowledge of the properties of the scattering medium responsible for the scintillations.  Determination of the absolute physical scale of the core shift does, however, require knowledge of the scintillation parameters, and we can only determine these quantities approximately.  The observed 520\,s time offset between the scintillations at the the lowest (4540\,MHz) and highest (9960\,MHz) frequency bands of our observations implies an angular separation of $\gtrsim 12\,\mu$as, for an assumed scattering screen distance of 10\,pc.  This translates to a physical scale of $\sim 0.10$\,pc at the source.  This physical scale probed is an order of magnitude smaller than typical core shifts obtained with VLBI measurements.  We further note that this technique easily detects the core-shift between frequency pairs separated by only $\sim 300\,$MHz, thus providing sub-microarcsecond resolution of the jet structure.


A major conclusion arising from our analysis is that that the often assumed frequency scaling of core position, $r_{\rm core} \propto \nu^{-1}$ \citep[e.\,g.][]{pushkarev12}, may not be applicable to all sources, in line with similar findings of \citet{lobanov98} and \citet{kudryavtseva11}. Further, our results hint at a physical difference between persistent intra-day variable (IDV) sources, and the broader population of AGN.

To place these results in a physical context, we have explored a simple model based on a hydrostatically confined jet traversing a pressure gradient.  The pressure profile implied by this model is steep: $p \propto r^{-n_p}; n_p \gtrsim 7$.  Such steep pressure gradients have previously been suggested in VLBI studies of the frequency dependent core shifts in AGN \citep{lobanov98}.




The discrepancy between the frequency scaling of core position observed in PKS\,1257--326, and the typical $r_{\rm core} \propto \nu^{-1}$ scaling observed in sources studied with VLBI may arise because different types of source lend themselves to different types of analysis. Only the largest core shifts can be detected with VLBI imaging. In an analysis of 277 objects by \citet{kovalev08}, only 10$\%$ gave reliable core shift measurements. The sources used in the detailed, multi-frequency VLBI study of the core shift effect by \citet{sokolovsky11} were selected based on their known large core shift, in addition to their bright optically thin jet features that were required to enable accurate registration of the images. 

More specifically, we suggest that the peculiar nature of the core-shift frequency dependence in PKS\,1257--326 is related to a number of other remarkable source properties.  The implied brightness temperature of the scintillating component in PKS\,1257-326 is high, $> 10^{12}\,$K (Bignall et al.\,2006), and the source has exhibited such bright emission at least as long as IDV has been observed, since 1995.  The persistence of this IDV over more than 15 years is relatively rare amongst IDV sources.  During the 4-epoch, one year duration MASIV survey, Lovell et al.\,(2008) found that sources which consistently exhibited IDV over all four epochs accounted for only 20\% of all IDV sources observed.   Moreover, the stability of the annual cycle in the source variability timescale (Bignall et al.\,2003, 2006) suggests that the size of the source is remarkably stable on a timescale of years.  Physically, this implies that the ultracompact jet in this source is both remarkably bright and stable.   It is tempting to speculate, therefore, that the hydrostatic confinement provided by the strong pressure gradient suggested by our simple model is responsible for the observed stability of the jet.


\acknowledgments
The Australia Telescope is funded by the Commonwealth of Australia for operation as a National Facility managed by CSIRO. The observations presented here were made by JAH as part of the CSIRO Astronomy \& Space Science (CASS) Vacation Scholarship Program. JAH thanks Dominic Schnitzeler for assistance with the observing setup.  Parts of this research were conducted by the Australian Research Council Centre of Excellence for All-sky Astrophysics (CAASTRO), through project number CE110001020.

{\it Facilities:} \facility{ATCA ()}

\appendix

\section{Deriving source structure from the time delay between scintillation lightcurves} \label{app:Correlation}

Here we relate the time delay observed between the scintillations at two frequencies to the angular displacement $\btheta$ between the centroids of the source emission at the two frequencies.  The time delay depends not only on the velocity, but also on the axial ratio, $R$, and orientation of the interstellar scintillation pattern.  

For intensity fluctuations $\Delta I({\bf r},t)$ measured at two locations ${\bf r}'$ and ${\bf r}' + {\bf r}$ on the observer's plane at time $t$, the intensity correlation function is defined as, 
\begin{eqnarray}
\rho({\bf r},t) = \frac{\langle \Delta I({\bf r}',t') \Delta I({\bf r}'+{\bf r},t'+t) \rangle }{\langle \Delta I^2 \rangle} .
\end{eqnarray}
We follow the treatment of Coles \& Kaufman (1978), in which the simplest approximation is that the contours of equal spatial correlation comprise a family of similar ellipses, and the model for the spatial correlation function takes the form
\begin{eqnarray}
\rho({\bf r},0)= f \left( \frac{| C {\bf r} |^2}{\sigma^2} \right), \label{rhoform}
\end{eqnarray}
where $f$ is a monotonically decreasing function of $|C {\bf r}|^2$.  If the scintles are elliptical with axial ratio $R$ and $x$ axis is inclined at an angle $\beta$ with respect to the major axis, then $C$ takes the form
\begin{eqnarray}
C = \left[  \begin{array}{cc} 
\cos \beta/\sqrt{R} & \sin \beta/\sqrt{R} \\
- \sqrt{R} \sin \beta & \sqrt{R} \cos \beta \\
\end{array} \right]. \label{Cform}
\end{eqnarray}

The scintillation pattern is assumed frozen onto a screen that moves past the observer at velocity ${\bf v}$, so there is a direction relation between the spatial and temporal dependence of the autcorrelation, \begin{eqnarray}
\rho({\bf r},t) = f \left( \frac{| C ({\bf r}-{\bf v} t) |^2}{\sigma^2} \right).
\end{eqnarray}
Surfaces of constant correlation correspond to curves of constant $| C ({\bf r}-{\bf v} t) |^2$.
The maximum of $\rho({\bf r},t)$ occurs at a time lag $\Delta t$ given by
\begin{eqnarray}
\Delta t = {\bf a} \cdot {\bf r} \label{ColesSoln}
\end{eqnarray}
where
\begin{eqnarray}
{\bf a} = \frac{(C^T C) \cdot {\bf v}}{| C \cdot {\bf v}|^2}. \label{aDefn}
\end{eqnarray}
Equations (\ref{rhoform}) $-$ (\ref{aDefn}) are derived in Coles \& Kaufman (1978), and are shown here for completeness.  

Now suppose a source possesses similar structure at two frequencies $\nu_1$ and $\nu_2$, but that they are displaced by an angle $\btheta$.  Then the scintilation fluctuations at $\nu_1$, $\Delta I_1({\bf r},t)$, are identical to those at $\nu_2$, $\Delta I_2({\bf r},t)$, except that they are displaced by a linear scale $D\btheta$ (see, e.g. Little \& Hewish 1966):
\begin{eqnarray}
I_2({\bf r}) = I_1({\bf r} - D \btheta),
\end{eqnarray}
where $D$ is the scattering screen distance.
The cross-correlation between $I_1$ and $I_2$ takes the form 
\begin{eqnarray}
\frac{\langle \Delta I_1 ({\bf r}',t') \Delta I_2({\bf r}'+{\bf r},t'+t) \rangle}{\langle \Delta I_1 \Delta I_2 \rangle} 
&=& \frac{\langle \Delta I_1 ({\bf r}',t') \Delta I_1({\bf r}'+{\bf r}-D \btheta,t'+t) \rangle}{\langle \Delta I_1 \Delta I_2 \rangle} \equiv \rho_{I_1 I_2}({\bf r}-D\btheta,t). \nonumber \\
\end{eqnarray}
Once again, we assume that $\rho_{I_1 I_2}$ is a montonically decreasing function of $r$ that takes the form given by equation (\ref{rhoform}).  Now, if the scintillations are measured at identical locations, we are interested in the maximum of $\rho_{I_1 I_2} (-D \btheta,t)$. In this case the time delay measured between $I_1$ and $I_2$ is the same as that given by eq.\ (\ref{ColesSoln}) above, with ${\bf r}$ replaced by $-D \btheta$, namely,
\begin{eqnarray}
\Delta t = - D {\bf a} \cdot \btheta = - \frac{D \btheta \cdot {\bf v} + (R^2-1) (D \btheta \times \hat{\bf S})({\bf v} \times \hat{\bf S}) }{v^2 + (R^2-1)({\bf v} \times \hat{\bf S})^2}.
\end{eqnarray}
where $\hat{\bf S} = (\cos \beta,\sin \beta)$ is the direction along which the scintles are oriented.  The cross product of a vector with $\hat{\bf S}$ is the component of that vector that points orthogonal to the elongation axis.  For instance, ${\bf r} \times \hat{\bf S}$ is the component of ${\bf r}$ orthogonal to the long axis of the scintillation pattern.

Finally, we note that the timescale of the scintillations can also be derived from the foregoing formalism, and is given by,
\begin{eqnarray}
t_{\rm scint} = \frac{\sigma \sqrt{R}}{\sqrt{v^2 + (R^2 -1) ({\bf v} \times \hat{\bf S})^2}}, \label{tscintEq}
\end{eqnarray}
where $\sigma$, defined by eq.\,(\ref{rhoform}), is the overall scale factor of the scintillation pattern.

\section{Core shift in the presence of a pressure gradient} \label{app:hydrostatic_confinement}

As discussed by \citet[][]{lobanov98}, the frequency dependence of core position may be influenced by the pressure gradients in the external medium if hydrostatic confinement is important. \citet[][]{lobanov98} plotted the frequency dependence of core position as a function of the power-law index of the pressure profile, but neglected the effect of the changing Doppler factor along the jet. Accordingly, we present a derivation of the properties of the core (diameter, distance along the jet, and spectral index) for a simple model of a hydrostatically confined jet in the presence of a power-law pressure profile, accounting for the effect of a radially dependent Doppler factor. 

Let $L$ be the path length through the source, $d$ the diameter of the jet perpendicular to the jet axis, $D_A$ the angular size distance, $\phi$ the jet viewing angle, and $\alpha^\prime_{\nu^\prime}$ the absorption coefficient in the source co-moving frame, at the rest frequency $\nu^{\prime} = \frac{1+z}{\delta} \nu$. Assuming the jet opening angle is small compared to the jet viewing angle (that is, we approximate the jet as a cylinder), we can approximate the optical depth as,
\begin{eqnarray}
\tau_\nu = L \alpha_\nu &\approx&  \frac{d}{\sin \phi} \alpha_\nu \nonumber \\
&=& \frac{d}{\sin \phi}  \frac{(1+z)}{\delta} \alpha^{\prime}_{\nu^\prime},
\end{eqnarray}
since $\nu \alpha _\nu$ is a relativistic invariant. For a power law electron distribution, the absorption coefficient is,
\begin{equation*}
\alpha^\prime_{\nu^\prime} = C(\alpha) k_e B^{\alpha + 1.5} \left( \frac{1+z}{\delta} \nu \right)^{-(\alpha + 2.5)} ,
\end{equation*}
where $C(\alpha)$ is a constant which depends only on the optically thin spectral index $\alpha$. So now,
\begin{equation}
\tau_\nu = \frac{d}{\sin \phi} C(\alpha) k_e B^{\alpha + 1.5} \left( \frac{1+z}{\delta} \right)^{-(\alpha + 1.5)} \nu^{-(\alpha + 2.5)}.
\end{equation}
Hence, at the $\tau_\nu = 1$ surface, which we identify with the scintillating component, or core: 
\begin{equation} \label{eqn:r_ratio}
\left( \frac{\nu}{\nu_0} \right)^{(\alpha + 2.5)} =  \frac{d(r)}{d(r_0)} \frac{k_e(r)}{k_e(r_0)} \left[ \frac{\delta(r) B(r)}{\delta(r_0) B(r_0)} \right]^{1.5 + \alpha}, 
\end{equation}
where $r_0$ is the position of the $\tau_\nu = 1$ surface, or core, at frequency $\nu_0$.

Hydrostatic confinement implies that the pressure inside the jet adjusts to the pressure of the surrounding medium. In that case, the diameter of the jet, $d(r)$, is determined entirely by the pressure gradient of the external medium, and the lateral expansion of the jet dictates the run of magnetic field, particle density, and Lorentz factor along the jet. Consider a jet with ultra-relativistic equation of state, and relativistic particle energy distribution of the form $N(\gamma) = K_e \gamma^{-a}$ between some minimum and maximum Lorentz factor, $\gamma_{\rm min}$ and $\gamma_{\rm max}$, confined by an ambient medium with pressure $p_{\rm ext} \propto r^{-n_p}$. In that case, conservation equations imply that the following relations hold:
\begin{eqnarray} \label{eqn:relationships2}
d(r) &\propto& r^{n_p/4} \\ \nonumber 
\Gamma(r) &\propto& r^{n_p/4} \\ \nonumber 
B_{||} &\propto& r^{-n_p/2} \\ \nonumber 
B_{\perp} &\propto& r^{-n_p/4} \\ \nonumber 
n &\propto& r^{-3n_p/4} \\ \nonumber 
\gamma_{\rm min} &\propto& r^{-n_p/4} \\ \nonumber 
k_e &\propto& r^{-\frac{n_p (\alpha+1.5)}{2}}  \nonumber 
\end{eqnarray}
Following \citet{konigl81} and \citet{lobanov98}, let us define $k_r$ such that
\begin{equation}
r_{\rm core} \propto \nu^{-1/k_r}. 
\end{equation}
and $m_B$ such that
\begin{equation}
B \propto r^{-\frac{m_B n_p}{4}}.
\end{equation}
In that case, $m_B = 1$ corresponds to a predominantly toroidal (perpendicular) magnetic field, while $m_B = 2$ corresponds to a predominantly poloidal (parallel) magnetic field. 
Approximating the radial dependence in Doppler factor as a power law of the form $\delta \propto r^{n_\delta}$, Equations \ref{eqn:r_ratio} and \ref{eqn:relationships2} give
\begin{equation} \label{eqn:k_r}
k_r = \frac{(n_p/4)[(1.5+\alpha)(2+m_B) - 1]  - (1.5 + \alpha) n_\delta}{2.5 + \alpha}.
\end{equation}
Note that $n_\delta$ may be positive or negative depending on the initial Lorentz factor, $\Gamma(r_0)$, and jet viewing angle $\phi$, and therefore the effect of the Doppler factor may be to steepen or flatten the relationship between r and $\nu$.

For a self-absorbed synchrotron source, the flux density at the peak frequency (which we associate with the flux density at the observed frequency) is,
\begin{eqnarray}
\label{eqn:S_m_general}
S_m &\propto& \nu_m^{5/2} ~ \theta_d^2 ~ B^{-1/2} ~ \delta^{1/2} . 
\end{eqnarray}
This expression, when combined with the functions d(r), B(r),  $\delta$(r) and r($\nu$), allows a prediction of the core spectral index \citep[see][equation 12]{konigl81}. Here we define the spectral index of the core, $\alpha_{\rm core}$, such that $S_\nu \propto \nu^{\alpha_{\rm core}}$, and
\begin{equation} \label{eqn:alpha_core}
\alpha_{\rm core} = \frac{5}{2} - \frac{1}{2 k_r} \left[  n_p\left( 1 + \frac{m_B}{4}  \right)  + n_\delta \right] .
\end{equation}


\clearpage

\end{document}